\documentclass[12pt]{article}
\usepackage{latexsym,graphicx}
\usepackage{subfigure}
\usepackage{amssymb}
\usepackage{amsmath}
\usepackage{amscd}
\usepackage{amsthm}
\usepackage[left=2cm,top=2.5cm,right=2.5cm,bottom=1.5cm]{geometry}          
\begin{document}
\begin{center}
\large{\bf{Deformation Retract and Folding of the $5D$ Schwarzchild Field}} \\
\vspace{10mm}
\normalsize{Nasr Ahmed$^a$,$^b$ and H. Rafat$^a$,$^c$}\\
\vspace{5mm}
\normalsize{$^{a}$ Mathematics Department, Faculty of Science, Taibah university, Saudi Arabia.\\$^b$Astronomy Department, National Research Institute of Astronomy and Geophysics, Helwan, Cairo, Egypt\footnote{abualansar@gmail.com}} \\
\normalsize{$^{c}$Department of Mathematics, Faculty of Science, Tanta University, Tanta, Egypt.\footnote{hishamrafat2000@yahoo.com}} \\
\vspace{2mm}
\end{center}
\vspace{5mm}
\begin{abstract}
In this article we introduce some types of the deformtion retracts of the $5D$ Schwarzchild space making use of Lagrangian equations. The retraction of this space into itself and into geodesics has been presented. The relation between folding and deformation retract of this space has been achieved. A relation for energy conservation similar to the one obtained in four dimensions has been obtained for the five dimensional case. 
\end{abstract}

\section{Introduction}

The real revolution in mathematical physics in the second half of twentieth century (and in pure mathematics itself) was algebraic topology and algebraic geometry \cite{topo}. In the nineteenth century, Mathematical physics was essentially the classical theory of ordinary and partial differential equations. The variational calculus, as a basic tool for physicists in theoretical mechanics, was seen with great reservation by mathematicians until Hilbert set up its rigorous foundation by pushing forward functional analysis. This marked the transition into the first half of twentieth century, where under the influence of quantum mechanics and relativity, mathematical physics turned mainly into functional analysis, complemented by the theory of Lie groups and by tensor analysis. All branches of theoretical physics still can expect the strongest impacts from use of the unprecedented wealth of results of algebraic topology and algebraic geometry of the second half of twentieth century \cite{topo}.\\

Today, the concepts and methods of topology and geometry have become an indispensable part of theoretical physics. They have led to a deeper understanding of many crucial aspects in condensed matter physics, cosmology, gravity, and particle physics. Moreover, several intriguing connections between only apparently disconnected phenomena have been revealed based on these mathematical tools \cite{geo}.\\
 
Topology enters General Relativity through the fundamental assumption that spacetime exists and is organized as a manifold. This means that spacetime has a well-defined dimension, but it also carries with it the inherent possibility of modified patterns of global connectivity, such as distinguish a sphere from a plane, or a torus from a surface of higher genus. Such modifications can be present in the spatial topology without affecting the time direction, but they can also have a genuinely spacetime character in which case the spatial topology changes
with time \cite{topo1}. The topology change in classical general relativity has been discussed in \cite{topo2}. See \cite{topo3} for some applications of differential topology in general relativity.\\

Most of the studies on deformation retract and folding, if not all, are pure mathematical studies. The authors believe that these two concepts  should be given more attention in modern mathematical physics. A topological studies of some famous metrics of mathematical physics could be a nice topological exploration start.

\subsection{Deformation Retract - Definitions}

The theory of deformation retract is very interesting topic in Euclidean and non-Euclidean spaces. It has been investigated from different points of view in many branches of topology and differential geometry. A retraction is a continuous mapping from the entire space into a subspace which preserves the position of all points in that subspace \cite{retra}.\\

(i) Let $M$ and $N$ be two smooth manifolds of dimensions $m$ and $n$ respectively. A map $f$ :$M\rightarrow N$ is said to be an isometric folding of $M$ into $N$ if and only if for every piecewise geodesic path $\gamma:J\rightarrow M$, the induced path $f o \gamma:J\rightarrow N$ is a piecewise geodesic and of the same
length as $\gamma$ \cite{robertson}. If $f$ does not preserve the lengths, it is called topological folding. Many types
of foldings are discussed in \cite{{3},{4},{6},{9}}. Some applications are discussed in \cite{{2},{10}}.\\

(ii) A subset $A$ of a topological space $X$ is called a retract of $X$, if there exists a continuous
map $r : X \rightarrow A $ such that\cite{11} \\
(a) $X$ is open \\
(b) $r(a) = a$, $\forall a \in A$. \\

(iii) A subset $A$ of a topological space $X$ is said to be a deformation retract if there exists a
retraction $r: X \rightarrow A$, and a homotopy $f : X \times I \rightarrow X $ such that\cite{11} \newline
$f (x, 0) = x$, $\forall x\in X$,\newline
$f(x, 1) = r (x)$, $\forall r\in X$,\newline
$f(a, t) = a$, $\forall a\in A, t \in [0,1]$.\\

The deformation retract is a particular case of homotopy equivalence, two spaces are homotopy equivalent if and only if they are both deformation retracts of a single larger space. \\
Deformation retracts of Stein spaces has been studied in \cite{stien} .The deformation retract of the 4D Schwarzchild metric has been discussed in \cite{hisham} where it was found that the retraction of the Schwarzchild space is  spacetime geodesic. The retraction of Eguchi-Hanson space has been discussed in \cite{nasr} in this paper we are going to discuss the retraction for the five dimensional case.

\subsection{Schwarzchild metric in five dimensions}

For the Schwarzchild metric in $(n+1)$ dimensions we can write \cite{Ricci}
\begin{equation}
ds^{2}=-\left(1-\frac{\mu}{r^{n-2}}\right)dt^{2}+\left(1-\frac{\mu}{r^{n-2}}\right)^{-1}dr^{2}+r^{2}d\Omega^{2}_{n-1}
\end{equation}

$d\Omega^{2}_{n-1}$ is defined as
\begin{equation}
d\Omega^{2}_{n-1}=d\chi^{2}_{2}+\sin^{2}\chi_{2}d\chi_{3}^{2}+...+\prod^{n-1}_{m=2} \sin^{2}\chi_{m}d\chi_{n}^{2} 
\end{equation}

In five dimensions $n=4$ and we have:

\begin{equation}\label{MV1}
ds^{2}=-\left(1-\frac{\mu}{r^{2}}\right)4\pi^{2}\mu d\tau^{2}+\left(1-\frac{\mu}{r^{2}}\right)^{-1}dr^{2}+r^{2}d\Omega^{2}_{3}
\end{equation}

Where the coordinate $t$ has been identified with a period $2\pi \sqrt{\mu} \tau$ to remove the singularity at $r^{2}=\mu$. If we set $\mu=2m$ then $dt^2=8\pi^{2}m d\tau^{2}$(see \cite{Ricci}).
The 5D flat metric can be written as 
\begin{equation}\label{flat}
ds^{2}=-dx_{o}^{2}+\sum_{i=1}^{4}dx_{i}^{2}
\end{equation}
Comparing (\ref{MV1}) and (\ref{flat}) making use of the basic metric definition $ds^2=g_{ij}dx^idx^j$. The following relations could be obtained 

\begin{eqnarray}
x_{o}=\pm\sqrt{\left(1-\frac{\mu}{r^{2}}\right)4\pi^{2}\mu \tau^{2}+C_{o}},~ x_{1}=\pm\sqrt{r^2+2r\sqrt{\mu}\tanh^{-1}\left(\frac{r}{\sqrt{\mu}}\right)-\mu \ln\left(1-\frac{r^2}{\mu}\right)+C_{1}},\nonumber \\ x_{2}=\pm\sqrt{r^{2}\theta^{2}+C_{2}},~x_{3}=\pm\sqrt{r^{2}\phi^{2}\sin^{2}\theta+C_{3}},~x_{4}=\pm\sqrt{r^{2}l^{2}\sin^{2}\theta \sin^{2}\phi+C_{4}} ~~~~~~~~~~~~~~~~~~
\end{eqnarray}

where $C_{o}$,$C_{1}$,$C_{2}$,$C_{3}$ and $C_{4}$ are constants of integration. These relations will play a key role in the study of the deformation retract of the 5D Schwarzchild spacetime.

\section{Euler-Lagrange equations of 5D Schwarzchild field}

In general relativity, the geodesic equation is equivalent to the Euler-Lagrange equations 
\begin{equation}
\frac{d}{d\lambda}\left(\frac{\partial L}{\partial \dot{x}^{\alpha}}\right)-\frac{\partial L}{\partial x^{\alpha}} =0, ~~~~i=1,2,3,4
\end{equation}
associated to the Lagrangian 
\begin{equation}
L(x^{\mu},\dot{x}^{\mu})=\frac{1}{2}g_{\mu \nu}\dot{x}^{\mu}\dot{x}^{\nu}
\end{equation}

The Lagrangian of the $5D$ Schwarzchild field can be written as

\begin{equation}
L=-\left(1-\frac{\mu}{r^{2}}\right)4\pi^{2}\mu \dot{\tau}^{2}+\left(1-\frac{\mu}{r^{2}}\right)^{-1}\dot{r}^{2}+r^{2}\left(\dot{\theta}^{2}+(\sin^{2}\theta) \dot{\phi}^{2}+(\sin^{2}\theta \sin^{2}\phi) \dot{l}^{2}\right)
\end{equation}
No explicit dependence on either $t$ or $l$, and thus $\frac{\partial L}{\partial \dot{t}}$ and $\frac{\partial L}{\partial \dot{l}}$ are constants of motion, i.e.
\begin{equation}
\left(1-\frac{\mu}{r^{2}}\right)2\pi\sqrt{\mu} \dot{\tau}=k ,~~~~~~~~~~~~r^{2}(\sin^{2}\theta \sin^{2}\phi) \dot{l}=h
\end{equation}

with $k$ and $h$ are constants. $h$ can be regarded as an equivalent of angular momentum per unit mass in the fifth dimension. using the Euler-Lagrange equations we get the full set of components as

\begin{eqnarray}
\frac{d}{d\lambda}\left(\frac{\dot{r}}{1-\frac{\mu}{r^{2}}}\right)-\left[-\frac{4\mu \pi^{2}}{r^{3}}\dot{\tau}^{2}-\frac{\mu \dot{r}^2}{r^{3}\left(1-\frac{\mu}{r^{2}}\right)^{2}}+r\left(\dot{\theta}^{2}+(\sin^{2}\theta) \dot{\phi}^{2}+(\sin^{2}\theta \sin^{2}\phi) \dot{l}^{2}\right)\right]=0\\ 
\frac{d}{d\lambda}(2r^{2}\dot{\theta})-r^{2}\sin2\theta(\dot{\phi}^{2}+\dot{l}^{2}\sin^{2}\phi)=0\\
\frac{d}{d\lambda}(2r^{2}\dot{\phi}\sin\theta)-r^{2}\dot{l}^{2}\sin\theta\sin2\phi=0\\
\frac{d}{d\lambda}\left(1-\frac{\mu}{r^{2}}\right)2\pi\sqrt{\mu} \dot{\tau}=0\\
\frac{d}{d\lambda}r^{2}\dot{l}\sin^{2}\theta \sin^{2}\phi=0.
\end{eqnarray} 

From $\left(1-\frac{\mu}{r^{2}}\right)2\pi\sqrt{\mu} \dot{\tau}=k$, setting $k=0$ gives two cases:
(1) $\dot{\tau}=0$ or $\tau=A$. If $A=0$ we get the following coordinates 

\begin{eqnarray}
x_{o}=\pm \sqrt{C_{o}},~x_{1}=\pm\sqrt{r^2+2r\sqrt{\mu}\tanh^{-1}\left(\frac{r}{\sqrt{\mu}}\right)-\mu \ln\left(1-\frac{r^2}{\mu}\right)+C_{1}},~~~~~~~~~~~~~~~~~~~~~~~~~~~~~\nonumber \\ 
x_{2}=\pm\sqrt{r^{2}\theta^{2}+C_{2}},~x_{3}=\pm\sqrt{r^{2}\phi^{2}\sin^{2}\theta+C_{3}},~x_{4}=\pm\sqrt{r^{2}l^{2}\sin^{2}\theta \sin^{2}\phi+C_{4}}~~~~~~~~~~~~~~~~~~~~~~\nonumber \\
x_{5}=\pm\sqrt{r^{2}\eta^{2}\sin^{2}\theta \sin^{2}\phi \sin^{2}l+C_{5}} ~~~~~~~~~~~~~~~~~~~~~~~~~~~~~~~~~~~~~~~~~~~~~~~~~~~~~~~~~~~~~~~~~
\end{eqnarray}

Since $x_{1}^{2}+x_{2}^{2}+x_{3}^{2}+x_{4}^{2}-x_{o}^{2}>0$ which is the great circle $S_{1}$ in the Schwarzchild
spacetime $S$. These geodesic is a retraction in Schwarzchild space. $ds^{2}>0$. $\mu=0$ is not allowed as it leads to undefined quantities.\\
From $r^{2}(\sin^{2}\theta\sin^{2}\phi) \dot{l}=h$, if $h=0$ we have three cases:(1) $\dot{l}=0$ or $l=B$. If $B=0$ we get the coordinates as:

\begin{eqnarray}
x_{o}=\pm\sqrt{\left(1-\frac{\mu}{r^{2}}\right)4\pi^{2}\mu \tau^{2}+C_{o}},~x_{1}=\pm\sqrt{r^2+2r\sqrt{\mu}\tanh^{-1}\left(\frac{r}{\sqrt{\mu}}\right)-\mu \ln\left(1-\frac{r^2}{\mu}\right)+C_{1}}\nonumber \\
x_{2}=\pm\sqrt{r^{2}\theta^{2}+C_{2}},~x_{3}=\pm\sqrt{r^{2}\phi^{2}\sin^{2}\theta+C_{3}},~x_{4}=\pm\sqrt{C_{4}}~~~~~~~~~~~~~~~~~~~~~~~~
\end{eqnarray}
This is the geodesic hyperspacetime $S_{3}$ of the Schwarzchild space $S$. This is
a retraction. $ds^{2}>0$.\\
(2) $\phi=0$ and in this case we get
\begin{eqnarray}
x_{o}=\pm\sqrt{\left(1-\frac{\mu}{r^{2}}\right)4\pi^{2}\mu \tau^{2}+C_{o}},~x_{1}=\pm\sqrt{r^2+2r\sqrt{\mu}\tanh^{-1}\left(\frac{r}{\sqrt{\mu}}\right)-\mu \ln\left(1-\frac{r^2}{\mu}\right)+C_{1}},\nonumber \\
x_{2}=\pm\sqrt{r^{2}\theta^{2}+C_{2}},~x_{3}=\pm\sqrt{C_{3}},~x_{4}=\pm\sqrt{C_{4}}.~~~~~~~~~~~~~~~~~~~~~~~~~ 
\end{eqnarray}
This is the geodesic hyperspacetime $S_{4}$ of the Schwarzchild space $S$. This is
a retraction. $ds^{2}>0$.\\
(3) $\theta=0$ and in this case we get
\begin{eqnarray}
x_{o}=\pm\sqrt{\left(1-\frac{\mu}{r^{2}}\right)4\pi^{2}\mu \tau^{2}+C_{o}},~x_{1}=\pm\sqrt{r^2+2r\sqrt{\mu}\tanh^{-1}\left(\frac{r}{\sqrt{\mu}}\right)-\mu \ln\left(1-\frac{r^2}{\mu}\right)+C_{1}}\nonumber \\
x_{2}=\pm\sqrt{C_{2}},~x_{3}=\pm\sqrt{C_{3}},~x_{4}=\pm\sqrt{C_{4}}.~~~~~~~~~~~~~~~~~~~~~~~~~~~~~
\end{eqnarray}

This is the geodesic hyperspacetime $S_{5}$ of the Schwarzchild space $S$. This is
a retraction. $ds^{2}>0$.\\

\textbf{Theorem1:}\\

The retraction of 5D Schwarzchild space is a 5D spacetime geodesic.
\section{deformation retract of Schwarzchild space}
The deformation retract of the $5D$ Schwarzchild space is defined as
\begin{equation}
\rho:S\times I \rightarrow S 
\end{equation}
where $S$ is the 5-dimensional Schwarzchild space and $I$ is the closed interval $[0,1]$. The retraction of $5D$ Schwarzchild space $S$
is defined as
\begin{equation}
R:S\rightarrow S_{1}, S_{2},S_{3},S_{4} ~and~ S_{5}.
\end{equation}
The deformation retract of the 5D Schwarzchild space $S$ into a geodesic $S_{1}\subset S$ is defined as 
\begin{eqnarray}
\rho(m,t)=(1-t)\left\{\pm\sqrt{\left(1-\frac{\mu}{r^{2}}\right)4\pi^{2}\mu \tau^{2}+C_{o}},  \right. ~~~~~~~~~~~~~~~~~~~~~~~~~~~~~~~~~~~~~~~~\\ \nonumber 
\left. \pm\sqrt{r^2+2r\sqrt{\mu}\tanh^{-1}\left(\frac{r}{\sqrt{\mu}}\right)-\mu \ln\left(1-\frac{r^2}{\mu}\right)+C_{1}},~\pm\sqrt{r^{2}\theta^{2}+C_{2}}, \right. \\ \nonumber
\left. \pm\sqrt{r^{2}\phi^{2}\sin^{2}\theta+C_{3}},~\pm\sqrt{r^{2}l^{2}\sin^{2}\theta \sin^{2}\phi+C_{4}} \right\} +t\left\{\pm\sqrt{C_{o}}\right. ~~~~~~~~~~~\\ \nonumber
\left.,~\pm\sqrt{r^2+2r\sqrt{\mu}\tanh^{-1}\left(\frac{r}{\sqrt{\mu}}\right)-\mu \ln\left(1-\frac{r^2}{\mu}\right)+C_{1}}~,~\pm\sqrt{r^{2}\theta^{2}+C_{2}}\right. \\ \nonumber
\left.,~\pm\sqrt{r^{2}\phi^{2}\sin^{2}\theta+C_{3}}~,~\pm\sqrt{r^{2}l^{2}\sin^{2}\theta \sin^{2}\phi+C_{4}} \right\}~~~~~~~~~~~~~~~~~~~~~~~~~
\end{eqnarray}
where
\begin{eqnarray}
\rho(m,0)=\left\{\pm\sqrt{\left(1-\frac{\mu}{r^{2}}\right)4\pi^{2}\mu \tau^{2}+C_{o}},  \right. ~~~~~~~~~~~~~~~~~~~~~~~~~~~~~~~~~~~~~~~~~~~~~~~~~~~\\ \nonumber 
\left. \pm\sqrt{r^2+2r\sqrt{\mu}\tanh^{-1}\left(\frac{r}{\sqrt{\mu}}\right)-\mu \ln\left(1-\frac{r^2}{\mu}\right)+C_{1}},~\pm\sqrt{r^{2}\theta^{2}+C_{2}}, \right. \\ \nonumber
\left. \pm\sqrt{r^{2}\phi^{2}\sin^{2}\theta+C_{3}},~\pm\sqrt{r^{2}l^{2}\sin^{2}\theta \sin^{2}\phi+C_{4}} \right\}~~~~~~~~~~~~~~~~~~~~~~~~~~~
\end{eqnarray}
,
\begin{eqnarray}
\rho(m,1)=\left\{\pm\sqrt{C_{o}}~,~\pm\sqrt{r^2+2r\sqrt{\mu}\tanh^{-1}\left(\frac{r}{\sqrt{\mu}}\right)-\mu \ln\left(1-\frac{r^2}{\mu}\right)+C_{1}}\right. \\ \nonumber
\left.,~\pm\sqrt{r^{2}\theta^{2}+C_{2}}~,~\pm\sqrt{r^{2}\phi^{2}\sin^{2}\theta+C_{3}}~,~\pm\sqrt{r^{2}l^{2}\sin^{2}\theta \sin^{2}\phi+C_{4}} \right\}
\end{eqnarray}

The deformation retract of the 5D Schwarzchild space $S$ into a geodesic $S_{2}\subset S$ is defined as 
\begin{eqnarray}
\rho(m,t)=(1-t)\left\{\pm\sqrt{\left(1-\frac{\mu}{r^{2}}\right)4\pi^{2}\mu \tau^{2}+C_{o}},\right. ~~~~~~~~~~~~~~~~~~~~~~~~~~~~~~~~~~~~~~~~~~~~~~~~~~\\ \nonumber 
\left. \pm\sqrt{r^2+2r\sqrt{\mu}\tanh^{-1}\left(\frac{r}{\sqrt{\mu}}\right)-\mu \ln\left(1-\frac{r^2}{\mu}\right)+C_{1}},~\pm\sqrt{r^{2}\theta^{2}+C_{2}}, \right. ~~~~~~~~~\\ \nonumber
\left. \pm\sqrt{r^{2}\phi^{2}\sin^{2}\theta+C_{3}}~,~\pm\sqrt{r^{2}l^{2}\sin^{2}\theta \sin^{2}\phi+C_{4}} \right\}+t\left\{\pm\sqrt{\left(1-\frac{\mu}{r^{2}}\right)4\pi^{2}\mu \tau^{2}+C_{o}}\right.\\ \nonumber 
\left.,~\pm\sqrt{r^2+2r\sqrt{\mu}\tanh^{-1}\left(\frac{r}{\sqrt{\mu}}\right)-\mu \ln\left(1-\frac{r^2}{\mu}\right)+C_{1}},~\pm\sqrt{r^{2}\theta^{2}+C_{2}}\right. ~~~~~~~~~~~~~~~~~\\ \nonumber
\left.,~\pm\sqrt{r^{2}\phi^{2}\sin^{2}\theta+C_{3}}~,~\pm\sqrt{C_{4}} \right\}~~~~~~~~~~~~~~~~~~~~~~~~~~~~~~~~~~~~~~~~~~~~~~~~~~~~~~~~~~~~~~~~~
\end{eqnarray}

The deformation retract of the 5D Schwarzchild space $S$ into a geodesic $S_{3}\subset S$ is defined as 
\begin{eqnarray}
\rho(m,t)=(1-t)\left\{\pm\sqrt{\left(1-\frac{\mu}{r^{2}}\right)4\pi^{2}\mu \tau^{2}+C_{o}},\right. ~~~~~~~~~~~~~~~~~~~~~~~~~~~~~~~~~~~~~~~~~~~~~~~~~~~~~~\\ \nonumber 
\left. \pm\sqrt{r^2+2r\sqrt{\mu}\tanh^{-1}\left(\frac{r}{\sqrt{\mu}}\right)-\mu \ln\left(1-\frac{r^2}{\mu}\right)+C_{1}},~\pm\sqrt{r^{2}\theta^{2}+C_{2}}, \right. ~~~~~~~~~~~~~\\ \nonumber
\left. \pm\sqrt{r^{2}\phi^{2}\sin^{2}\theta+C_{3}}~,~\pm\sqrt{r^{2}l^{2}\sin^{2}\theta \sin^{2}\phi+C_{4}} \right\}+t\left\{\pm\sqrt{\left(1-\frac{\mu}{r^{2}}\right)4\pi^{2}\mu \tau^{2}+C_{o}}~,~\right. \\ \nonumber
\left.,~\pm\sqrt{r^2+2r\sqrt{\mu}\tanh^{-1}\left(\frac{r}{\sqrt{\mu}}\right)-\mu \ln\left(1-\frac{r^2}{\mu}\right)+C_{1}}~,~\pm\sqrt{r^{2}\theta^{2}+C_{2}}\right.~~~~~~~~~~~~~~~~~~~~ \\ \nonumber
\left.,~\pm\sqrt{C_{3}}~,~\pm\sqrt{C_{4}} \right\} ~~~~~~~~~~~~~~~~~~~~~~~~~~~~~~~~~~~~~~~~~~~~~~~~~~~~~~~~~~~~~~~~~~~~~~~~~~~~~~~~~~~~~~
\end{eqnarray}
The deformation retract of the 5D Schwarzchild space $S$ into a geodesic $S_{4}\subset S$ is defined as 
\begin{eqnarray}
\rho(m,t)=(1-t)\left\{\pm\sqrt{\left(1-\frac{\mu}{r^{2}}\right)4\pi^{2}\mu \tau^{2}+C_{o}},\right.~~~~~~~~~~~~~~~~~~~~~~~~~~~~~~~~~~~~~~~~~~~~~~~~~~~~~~~ \\ \nonumber 
\left. \pm\sqrt{r^2+2r\sqrt{\mu}\tanh^{-1}\left(\frac{r}{\sqrt{\mu}}\right)-\mu \ln\left(1-\frac{r^2}{\mu}\right)+C_{1}},~\pm\sqrt{r^{2}\theta^{2}+C_{2}}, \right. ~~~~~~~~~~~~~~~~\\ \nonumber
\left. \pm\sqrt{r^{2}\phi^{2}\sin^{2}\theta+C_{3}}~,~\pm\sqrt{r^{2}l^{2}\sin^{2}\theta \sin^{2}\phi+C_{4}} \right\}+t\left\{\pm\sqrt{\left(1-\frac{\mu}{r^{2}}\right)4\pi^{2}\mu \tau^{2}+C_{o}}\right. \\ \nonumber
\left.,~\pm\sqrt{r^2+2r\sqrt{\mu}\tanh^{-1}\left(\frac{r}{\sqrt{\mu}}\right)-\mu \ln\left(1-\frac{r^2}{\mu}\right)+C_{1}}~,~\pm\sqrt{C_{2}}~,~\pm\sqrt{C_{3}}~,~\pm\sqrt{C_{4}}  \right\}
\end{eqnarray}
Now we are going to discuss the folding $f$ of the 5D Schwarzchild space $S$. Let $f:S\rightarrow S$ where
\begin{equation}
f(x_{1},x_{2},x_{3},x_{4},x_{5})=(x_{1},x_{2},x_{3},x_{4},\left|x_{5}\right|)
\end{equation}
An isometric folding of the 5D Schwarzchild space into itself may be defined as:
\begin{eqnarray}
f:\left\{\pm\sqrt{\left(1-\frac{\mu}{r^{2}}\right)4\pi^{2}\mu \tau^{2}+C_{o}},\right.~~~~~~~~~~~~~~~~~~~~~~~~~~~~~~~~~~~~~~~~~~~~~~~~~~~~~~~~~~~~~~~~~~~~ \\ \nonumber 
\left. \pm\sqrt{r^2+2r\sqrt{\mu}\tanh^{-1}\left(\frac{r}{\sqrt{\mu}}\right)-\mu \ln\left(1-\frac{r^2}{\mu}\right)+C_{1}},~\pm\sqrt{r^{2}\theta^{2}+C_{2}}, \right. ~~~~~~~~~~~~~~~~~~\\ \nonumber
\left. \pm\sqrt{r^{2}\phi^{2}\sin^{2}\theta+C_{3}}~,~\pm\sqrt{r^{2}l^{2}\sin^{2}\theta \sin^{2}\phi+C_{4}} \right\}\rightarrow \left\{\pm\sqrt{\left(1-\frac{\mu}{r^{2}}\right)4\pi^{2}\mu \tau^{2}+C_{o}},\right. ~\\ \nonumber 
\left. \pm\sqrt{r^2+2r\sqrt{\mu}\tanh^{-1}\left(\frac{r}{\sqrt{\mu}}\right)-\mu \ln\left(1-\frac{r^2}{\mu}\right)+C_{1}},~\pm\sqrt{r^{2}\theta^{2}+C_{2}}, \right. ~~~~~~~~~~~~~~~~~~\\ \nonumber
\left. \pm\sqrt{r^{2}\phi^{2}\sin^{2}\theta+C_{3}}~,~\left|\pm\sqrt{r^{2}l^{2}\sin^{2}\theta \sin^{2}\phi+C_{4}}\right| \right\}~~~~~~~~~~~~~~~~~~~~~~~~~~~~~~~~~~~~~~~~~~~
\end{eqnarray}
The deformation retract of the folded 5D Schwarzchild space $S$ into the folded $S_{1}$ is
\begin{eqnarray}
\rho f: \left\{\pm\sqrt{\left(1-\frac{\mu}{r^{2}}\right)4\pi^{2}\mu \tau^{2}+C_{o}},\right.~~~~~~~~~~~~~~~~~~~~~~~~~~~~~~~~~~~~~~~~~~~~~~~~~~~~~~~~~~~~~~~~~~~~~~ \\ \nonumber 
\left. \pm\sqrt{r^2+2r\sqrt{\mu}\tanh^{-1}\left(\frac{r}{\sqrt{\mu}}\right)-\mu \ln\left(1-\frac{r^2}{\mu}\right)+C_{1}},~\pm\sqrt{r^{2}\theta^{2}+C_{2}}, \right. ~~~~~~~~~~~~~~~~~~~~\\ \nonumber
\left. \pm\sqrt{r^{2}\phi^{2}\sin^{2}\theta+C_{3}}~,~\left|\pm\sqrt{r^{2}l^{2}\sin^{2}\theta \sin^{2}\phi+C_{4}}\right| \right\}\times I \rightarrow 
\left\{\pm\sqrt{\left(1-\frac{\mu}{r^{2}}\right)4\pi^{2}\mu \tau^{2}+C_{o}},\right. \\ \nonumber 
\left. \pm\sqrt{r^2+2r\sqrt{\mu}\tanh^{-1}\left(\frac{r}{\sqrt{\mu}}\right)-\mu \ln\left(1-\frac{r^2}{\mu}\right)+C_{1}},~\pm\sqrt{r^{2}\theta^{2}+C_{2}}, \right. ~~~~~~~~~~~~~~~~~~~~~~~~\\ \nonumber
\left. \pm\sqrt{r^{2}\phi^{2}\sin^{2}\theta+C_{3}}~,~\left|\pm\sqrt{r^{2}l^{2}\sin^{2}\theta \sin^{2}\phi+C_{4}}\right| \right\}~~~~~~~~~~~~~~~~~~~~~~~~~~~~~~~~~~~~~~~~~~~~~~~~~
\end{eqnarray}
with 
\begin{eqnarray}
\rho f(m,t)= (1-t)\left\{S\right\}+t\left\{S_{1}\right\}~~~~~~~~~~~~~~~~~~~~~~~~~~~~~~~~~~~~~~~~~~~~~~~~~~~~~~~~~~~~~~~~~~~~~~~\\ \nonumber
           =(1-t)\left\{\pm\sqrt{\left(1-\frac{\mu}{r^{2}}\right)4\pi^{2}\mu \tau^{2}+C_{o}},  \right. ~~~~~~~~~~~~~~~~~~~~~~~~~~~~~~~~~~~~~~~~~~~~~~~~~\\ \nonumber 
\left. \pm\sqrt{r^2+2r\sqrt{\mu}\tanh^{-1}\left(\frac{r}{\sqrt{\mu}}\right)-\mu \ln\left(1-\frac{r^2}{\mu}\right)+C_{1}},~\pm\sqrt{r^{2}\theta^{2}+C_{2}}, \right. ~~~~~~~~~\\ \nonumber
\left. \pm\sqrt{r^{2}\phi^{2}\sin^{2}\theta+C_{3}},~\pm\sqrt{r^{2}l^{2}\sin^{2}\theta \sin^{2}\phi+C_{4}} \right\} +t\left\{\pm\sqrt{C_{o}}\right. ~~~~~~~~~~~~~~~~~~~~\\ \nonumber
\left.,~\pm\sqrt{r^2+2r\sqrt{\mu}\tanh^{-1}\left(\frac{r}{\sqrt{\mu}}\right)-\mu \ln\left(1-\frac{r^2}{\mu}\right)+C_{1}}~,~\pm\sqrt{r^{2}\theta^{2}+C_{2}}\right.~~~~~~~~~ \\ \nonumber
\left.,~\pm\sqrt{r^{2}\phi^{2}\sin^{2}\theta+C_{3}}~,~\pm\sqrt{r^{2}l^{2}\sin^{2}\theta \sin^{2}\phi+C_{4}} \right\}~~~~~~~~~~~~~~~~~~~~~~~~~~~~~~~~~~~
\end{eqnarray}
The same for $S_{2}$,$S_{3}$,$S_{4}$ and $S_{5}$. Then the following theorem has been proved \\

\textbf{Theorem2:}\\
 
Under the defined folding, the deformation retract of the folded 5D Schwarzchild space into the folded geodesics is the same as the deformation retract of the 5D Schwarzchild space into the geodesics.\\

\textbf{Theorem3:}\\

 The end of the limits of the folding of $n-$dimensional Schwarzchild space $S^{n}$ is a $0-$dimensional geodesic, it is a minimal retraction.\\
 
\textbf{Proof:} \\

let $ f_{1}:S^{n}\rightarrow S^{n}$, $ f_{2}: f_{1}(S^{n}) \rightarrow f_{1}(S^{n})$ , $f_{3}: f_{2}(f_{1}(S^{n})) \rightarrow f_{2}(f_{1}(S^{n}))$\dots \\

$f_{n}: f_{n-1}(f_{n-2}(f_{n-3}(\dots (f_{1}(S^{n}))))) \rightarrow f_{n-1}(f_{n-2}(f_{n-3}(\dots (f_{1}(S^{n})))))$.\\

$\lim _{n\rightarrow \infty} f_{n-1}(f_{n-2}(f_{n-3}(\dots (f_{1}(S^{n}))))) = n-1$ dimensional Schwarzchild space $S^{n-1}$. \\

Also, let  \\

$ h_{1}:S^{n-1}\rightarrow S^{n-1}$, $ h_{2}: h_{1}(S^{n-1}) \rightarrow h_{1}(S^{n-1})$ , $h_{3}: h_{2}(h_{1}(S^{n-1})) \rightarrow h_{2}(h_{1}(S^{n-1}))$\dots \\

$h_{m}: h_{m-1}(h_{m-2}(h_{m-3}(\dots (h_{1}(S^{n-1}))))) \rightarrow h_{m-1}(h_{m-2}(h_{m-3}(\dots (h_{1}(S^{n-1})))))$.\\

$\lim _{n\rightarrow \infty}h_{m-1}(h_{m-2}(h_{m-3}(\dots (h_{1}(S^{n-1}))))) = n-2$ dimensional Schwarzchild space $S^{n-2}$. \\

Consequently,\\

$\lim_{S\rightarrow \infty}\lim_{m\rightarrow \infty}\lim_{n\rightarrow \infty}-K_{S}(h_{m}(f_{n}(S^{n})))=0-$dimensional Schwarzchild space.

\section{energy conservation relation for $5D$ Schwarzchild field}
For the case of $(\theta = \phi= \pi/2)$, the equations to be solved are
\begin{equation} \label{1}
\left(1-\frac{\mu}{r^{2}}\right)\dot{t}=k
\end{equation}

\begin{equation} \label{2}
c^{2}=-c^{2}\left(1-\frac{\mu}{r^{2}}\right)\dot{t}^{2}+\left(1-\frac{\mu}{r^{2}}\right)^{-1}\dot{r}^{2}+r^{2}\dot{l}^{2}
\end{equation}

\begin{equation} \label{3}
r^{2}\dot{l}=h
\end{equation}

making use of (\ref{1}) and (\ref{3}) in (\ref{2}) and after some manipulations we get

\begin{equation} \label{5DEC}
\frac{1}{2}\dot{r}^{2}+\frac{h^{2}}{2r^{2}}\left(1-\frac{\mu}{r^{2}}\right)+\frac{c^{2}\mu}{2r^{2}}=c^{2}(1-k^{2})
\end{equation}

For the potential in higher dimensions, we recall the familiar Newton law in $n+4$ dimensions \cite{extra}
\begin{equation}
V_{n+4}\simeq \frac{G_{n+4} M}{r_{n}^{n+1}}
\end{equation}
So, for 5 dimensions the potential is inversly proportional to the square of $r$:
\begin{equation}
V_{5}\simeq \frac{G_{5} M}{r^{2}}
\end{equation}
The term
\begin{equation}
\frac{h^{2}}{2r^{2}}\left(1-\frac{\mu}{r^{2}}\right)+\frac{c^{2}\mu}{2r^{2}}
\end{equation}
represents the 5D potential and the equation (\ref{5DEC}) is the Energy conservation formula in 5D. Equation $(\ref{5DEC})$ is the $5D$ analogy of the 4D one for the case of 4D Schwarzchild field

\begin{equation} 
\frac{1}{2}\dot{r}^{2}+\frac{h^{2}}{2r}\left(1-\frac{\mu}{r}\right)-\frac{c^{2}\mu}{2r}=c^{2}(k^{2}-1)
\end{equation}
which expresses energy conservation in 4D with potential 
\begin{equation}
V_{4}=\frac{h^{2}}{2r^{2}}\left(1-\frac{\mu}{r}\right)-\frac{c^{2}\mu}{2r}
\end{equation}
\section{Conclusion}

The deformation retract of the five dimensional Schwarzchild space has been investigated by making use of Lagrangian equations. The retraction of this space into itself and into geodesics has been presented. The deformation retraction of the five dimensional Schwarzchild space is five dimensional space-time geodesics which found to be a great circle. The folding of this space has been discussed and it was found that this folding, and any folding homeomorphic to that folding, have the same deformation retract of the five dimensional Schwarzchild space into a geodesic. The energy conservation relation for the studied five dimensional Schwarzchild metric has been derived and compared to the ordinary four dimensional one.

\end{document}